\documentclass[10pt, conference, letterpaper]{IEEEtran}
\IEEEoverridecommandlockouts
\usepackage{cite}
\usepackage{amsmath,amssymb,amsfonts}
\usepackage{algorithmic}
\usepackage{graphicx}
\usepackage{textcomp}
\usepackage[dvipsnames,svgnames]{xcolor}
\usepackage{pbalance}
\usepackage{siunitx}
\usepackage[switch]{lineno}
\usepackage[sort=none]{glossaries}
\usepackage{multirow}
\usepackage{booktabs}
\usepackage{tabularx}

\usepackage[absolute,showboxes]{textpos}

\usepackage{caption}
\usepackage[position=b]{subcaption}

\usepackage[hidelinks]{hyperref}

\newif\ifanonymous

\let\orgautoref\autoref
\renewcommand{\autoref}
{\def\sectionautorefname{Section}\def\subsectionautorefname{Section}\def\subsubsectionautorefname{Section}\def\figureautorefname{Fig.}\def\equationautorefname{Eq.}\orgautoref}

\usepackage{pifont}

\usepackage{xspace}
\newcommand{\etal}{\textit{et al.}~}
\newcommand{\eg}{\textit{e.g.,}~}
\newcommand{\ie}{\textit{i.e.,}~}

\newcommand{\one}{({\em i})\xspace}
\newcommand{\two}{({\em ii})\xspace}
\newcommand{\three}{({\em iii})\xspace}
\newcommand{\four}{({\em iv})\xspace}

\makeatletter
\renewcommand{\paragraph}[1]{\vspace*{0.03in}\noindent{\bf #1.}\hspace{0.25ex \@plus1ex \@minus.2ex}}
\makeatother

\makeatletter
\newcommand{\paragraphc}[1]{\vspace*{0.03in}\noindent{\bf #1}\hspace{1ex \@minus.2ex}}
\makeatother

 \IEEEoverridecommandlockouts

\setacronymstyle{long-short}
\newacronym
 {ml}{ML}{machine learning}

\newacronym
 {llrf}
 {LLRF}
 {Low-Level RF}

\newacronym
 {xfel}
 {Eu-XFEL}
 {European X-ray Free-Electron Laser}

\newacronym
 {ro}
 {RO}
 {Ring Oscillator}

\newacronym
  {pdm}
  {PDM}
  {Propagation Delay Measurement}

\newacronym
  {sgd}
  {SGD}
  {Stochastic Gradient Descent}

\newacronym
  {mape}
  {MAPE}
  {Mean Average Percentage Error}
 
\begin{document}

\bstctlcite{IEEEexample:BSTcontrol}

\setlength{\TPHorizModule}{\paperwidth}
\setlength{\TPVertModule}{\paperheight}
\TPMargin{5pt}
\begin{textblock}{0.8}(0.1,0.02)
     \noindent
     \footnotesize
     If you cite this paper, please use the DSD reference:
     Leandro Lanzieri, Lukasz Butkowski, Jiri Kral, Goerschwin Fey, Holger Schlarb, and Thomas C. Schmidt.
     Studying the Degradation of Propagation Delay on FPGAs at the European XFEL.
     In \emph{Proceedings of the 27th Euromicro Conference on Digital System Design (DSD)}, IEEE, 2024.
\end{textblock}

\title{Studying the Degradation of Propagation Delay on FPGAs at the European XFEL}

\ifanonymous
\author{\IEEEauthorblockN{Paper \#NNN, \pageref{lastpage}~pages (incl. references)}}
\else
\author{
    \IEEEauthorblockN{Leandro Lanzieri\IEEEauthorrefmark{1}\IEEEauthorrefmark{2}\IEEEauthorrefmark{3},
                      Lukasz Butkowski\IEEEauthorrefmark{1},
                      Jiri Kral\IEEEauthorrefmark{1},
                      Goerschwin Fey\IEEEauthorrefmark{2},
                      Holger Schlarb\IEEEauthorrefmark{1}, and
                      Thomas C. Schmidt\IEEEauthorrefmark{3} \\
    }

    \IEEEauthorblockA{
        \IEEEauthorrefmark{1}Deutsches Elektronen-Synchrotron DESY, Germany \\
        \{leandro.lanzieri, lukasz.butkowski, jiri.kral, holger.schlarb\}@desy.de
    }

    \IEEEauthorblockA{
        \IEEEauthorrefmark{2}Hamburg University of Technology, Germany $\cdot$ goerschwin.fey@tuhh.de
    }

    \IEEEauthorblockA{
        \IEEEauthorrefmark{3}Hamburg University of Applied Sciences, Germany $\cdot$ t.schmidt@haw-hamburg.de
    }
}
\fi

\maketitle

\begin{abstract}
An increasing number of unhardened commercial-off-the-shelf embedded devices are deployed under harsh operating conditions and in highly-dependable systems.
Due to the mechanisms of hardware degradation that affect these devices, ageing detection and monitoring are crucial to prevent critical failures.
In this paper, we empirically study the propagation delay of 298 naturally-aged FPGA devices that are deployed in the European XFEL particle accelerator.
Based on in-field measurements, we find that operational devices show significantly slower switching frequencies than unused chips, and that increased gamma and neutron radiation doses correlate with increased hardware degradation.
Furthermore, we demonstrate the feasibility of developing machine learning models that estimate the switching frequencies of the devices based on historical and environmental data.
\end{abstract}


\section{Introduction} \label{sec:introduction}

The \gls{xfel} \cite{xfel-06} is a linear hard X-ray accelerator that energizes electrons up to \si{\qty{17.5}{\giga\electronvolt}}.
This is achieved by means of \si{\num{768}} superconducting radio frequency (RF) cavities placed along \si{\qty{1.7}{\kilo\meter}} of tunnel.
The \gls{llrf} system \cite{llrf-baag-12} oversees the precise control of the applied RF signals, and is composed of multiple interconnected embedded systems deployed on MicroTCA crates.
Unlike in other facilities of its kind, the electronic components in the \gls{xfel} are placed inside the accelerator tunnel to avoid communication latency, thus complying with the strict timing requirements of the operation.
In order to reduce cost on this large-scale deployment, the electronic devices that constitute the \gls{llrf} system are non-hardened Commercial-Off-The-Shelf (COTS) devices.
Due to the long deployment time and the enhanced gamma and neutron radiations to which the devices are exposed, there are concerns over the degradation effects on the hardware.
Indeed, ageing mechanisms such as total ionizing dose, bias temperature instability, and hot-carrier injection can degrade electronic components and induce system failures.

Harsh operating conditions accelerate degradation mechanisms on electronic components.
This reduces the operational lifespan of devices and increases the probability of failures.
The \gls{xfel} \gls{llrf} system is a critical and highly-dependable element of the accelerator operation.
Given the enormous cost that downtime entails for the facility, it is crucial to understand and monitor the effects of hardware degradation mechanisms \cite{lmfssw-tadmes-23}.

Environmental factors such as temperature \cite{nbti-s-07, rahd-swz-10} and radiation \cite{tidc-b-06} degrade silicon components.
Indeed, physical variations at transistor level affect parameters such as threshold voltages \cite{lkfss-aaesl-23} and maximum switching frequencies of electronic devices, thus impacting the overall performance of the system.
Degradation effects are particularly detrimental for high-speed real-time embedded applications, where modifications in the propagation times of signals inside the chip can produce timing faults with critical consequences.

\begin{figure}[t]
    \centering
    \includegraphics[width=\linewidth]{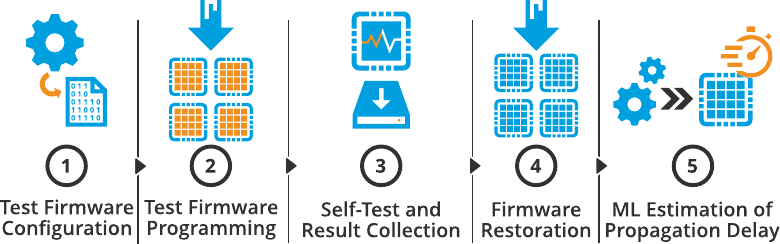}
    \setlength{\belowcaptionskip}{-10pt}
    \caption{Procedure to perform propagation delay measurements and estimations on deployed FPGA devices by means of an online non-concurrent self-test firmware.}
    \label{fig:procedure}
\end{figure}

The usage of \glspl{ro} to sense signal propagation delay of logical elements in FPGAs has been previously explored \cite{oshf-zh-10, dmma-sm-24}.
\glspl{ro} can also convey information such as temperature or voltage, serving as digital probes of physical magnitudes.
\glspl{ro} grant the advantage of low-resource requirements and high-precision placement on the die.

In this work, we study the degradation of the propagation delay on FPGA devices deployed in the \gls{xfel}.
In detail, our contributions are
\one  propagation delay measurements from \si{\num{298}} naturally-aged FPGAs, which have been deployed for seven years in the particle accelerator as part of the \gls{llrf} system,
\two a degradation study of the devices based on a comparison with unused FPGAs that have been kept in storage,
\three the analysis of the impact of variable radiations, such as gamma and neutron rays,
and \four the training and evaluation of various \gls{ml} models in frequency estimation of \glspl{ro} based on parameters such as FPGA die temperature, core voltage, and radiation doses.
To this end, we develop and apply an online self-test procedure (see \autoref{fig:procedure}) to measure the propagation delay of the hardware.
We aim to answer whether the enhanced radiation background has noticeably impacted the propagation delay of the FPGAs, and whether the propagation delay of operating devices substantially differs from those unused.

The remainder of the paper is structured as follows.
In \autoref{sec:background}, we introduce hardware degradation mechanisms at play in the studied deployment, explain the application of \glspl{ro} as on-chip sensors for FPGAs, and discuss relevant literature.
We then describe the utilized measurement procedure, the developed module, and the followed \gls{ml} training process in \autoref{sec:measurement}.
In \autoref{sec:results}, we present and analyse the measurement results and evaluate the performance of the \gls{ml} models.
Finally, we summarize our conclusions and propose future research steps in \autoref{sec:conclusions_future_work}.
 \section{Background and Related Work}\label{sec:background}

\subsection{Ageing Mechanisms on Transistors}\label{subsec:ageing_transistors}

Electronic components usually undergo degradation processes during operation, due to the gradual effects of ageing mechanisms on the hardware.
The time required for the ageing effects to manifest themselves highly depends on the environmental and operational conditions.
Transistors are susceptible to various silicon degradation mechanisms, which can be accelerated by environmental factors, causing performance reductions and affecting the overall system functionality.

In a constant effort to reduce the size of digital electronic components, transistor manufacturers have focused on developing technologies with smaller footprints.
Due to a non-proportional reduction of the applied voltage in relation to the device dimensions, smaller transistors are subjected to stronger electric fields, which can speed up degradation mechanisms.
Moreover, with the introduction of compact 3D Field Effect Transistors (FETs) such as FinFETs and gate-all-around FETs (GAAFETs), self heating susceptibility has become a performance threat \cite{ispr-vs-19}.
As a result of their constrained thermal paths to the ambient, these 3D structures can suffer an increase in operating temperature and in turn an acceleration of various ageing processes on the silicon.

\paragraph{Bias Temperature Instability} BTI is an ageing mechanism that affects transistors \cite{nbti-s-07}, causing an increment of their threshold voltage (\(V_{th}\)), and a degradation of parameters such as transconductance and drain current.
P-channel FETs are particularly impacted by Negative BTI (NBTI), while Positive BTI (PBTI) influences n-channel devices.
According to the Reaction-Diffusion model \cite{gdrm-os-95} that explains BTI, \(SiH\) bonds with low energies break on the silicon oxide-substrate interface (\(SiO_2/Si\)) under applied electric fields and elevated temperatures.
This causes positive charges to diffuse into the oxide, which leaves dangling bonds behind.
After enough positive charges accumulate in the gate oxide, an electric field appears opposing the activation one.
This induces a variation in \(V_{th}\) and reduces the maximum switching frequency.

\paragraph{Hot Carrier Injection} When carriers in the transistor are influenced by a high lateral electric field, they can gain enough kinetic energy to surpass the one of the silicon lattice \cite{rwmc-swvs-09}.
At this state, carriers are said to be \emph{"hot"}, and some of them have enough energy to diffuse into the gate oxide and damage the interface.
This effect is known as Hot Carrier Injection (HCI).
Similarly to BTI, HCI causes an increment of the transistor threshold voltage and degrades the device characteristics \cite{hcff-jlkk-16}.

\paragraph{Time-Dependent Dielectric Breakdown} Through the TDDB process, defects in the \(SiO_2\) layer of the transistor accumulate as a function of time.
This accumulation causes an increment of the gate leakage current \cite{iact-rlnk-17}, which can slow down the switching frequency of the transistor.
With a higher accumulation of defects in the oxide, a resistive path is formed.
This increases the gate current and operational temperature of the device.
When there is enough gate current, a thermal runaway eventually breaks the insulator and renders the transistor inoperable.

\paragraph{Total Ionizing Dose} Silicon devices can also be affected by radiation-induced trapped charges when the material is ionized as a result of interactions with high-energy photons or charged particles.
The Total Ionizing Dose (TID) represents the amount of energy that is transferred to the material during these interactions.
The accumulation of charges over time alters the transistor characteristics \cite{stro-sbwm-13}, potentially influencing its performance.

\begin{figure}[!t]
    \centering
    \includegraphics[width=\linewidth]{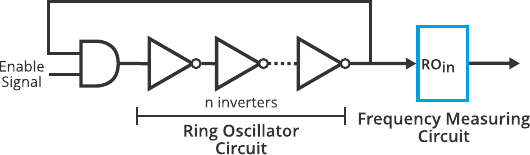}
    \caption{Diagram of a ring-oscillator-based propagation delay sensor.}
    \label{fig:ro_sensor}
\end{figure}

\subsection{Ring Oscillator Sensors on FPGAs}\label{subsec:ro_fpga}

As many other digital components, FPGAs are affected by ageing mechanisms that act on their transistor-based circuits.
Particularly, degradation processes such as BTI, HCI, and TID can slow the transistors down and increment the propagation times of signals.
Such delays can potentially produce timing faults and system malfunctions with critical consequences.

FPGAs are highly flexible devices due to their ability to be reprogrammed even after deployed.
This flexibility makes them great candidates to be configured with online on-chip testing circuits, which can potentially be removed during operation.
The \gls{ro} is a widely utilized characterization and degradation circuit for FPGAs \cite{lmfssw-tadmes-23}.

A typical sensor based on a \glspl{ro} is depicted in \autoref{fig:ro_sensor}.
The \gls{ro} is composed of an odd number of inverter gates connected such that they create a closed loop.
This forms an astable circuit that produces an oscillating digital signal.
At a given temperature and voltage, the frequency of the output signal \(f\) is given by \autoref{eq:ro-freq}, where \(n\) is the number of inverters and \(t_p\) their propagation time.

\begin{equation}
    f = \frac{1}{2 \cdot n \cdot t_p}
    \label{eq:ro-freq}
\end{equation}

Optionally, an enable signal can be added to switch the oscillation on and off.
Finally, a frequency measurement circuit (\eg a counter) is connected to the output signal.

Zick \etal \cite{oshf-zh-10} proposed a method to measure physical variations on FPGAs concurrently with the application.
The authors developed a compact sensor that consists of only 8 Look-Up Tables (LUTs) to achieve a high resource efficiency.
The \glspl{ro} were designed high sensitivity to temperature changes, in order to find high-temperature areas on the chip.
The authors were able to measure propagation times and estimate transistor current leakage profiles, power usage, and temperature.

\gls{ro}-based sensors have also been used to detect hardware degradation.
Li \etal \cite{iama-lhww-22} deployed \glspl{ro} on \si{\qty{28}{\nano\meter}} FPGAs to study the effects of NBTI under accelerated ageing conditions.
With the data collected from the measurements on artificially-aged devices, the authors trained various \gls{ml} models to estimate frequency degradation as a function of time, reaching a root mean squared error of \si{\qty{0.262}{\percent}}.
Although the degradation tests and predictions showed promising results, the methods were not evaluated on naturally-aged devices.
Instead, increased temperatures were applied to accelerate the ageing process of the FPGAs.

Pfeifer \etal developed a method based on Block RAMs to measure oscillations from \glspl{ro} called ``Reliability-on-Chip'' \cite{rloc-pkp-14}.
By undersampling the oscillations with the on-chip memories, a softcore is able to calculate propagation delays from signal streams.
A degradation evaluation on various devices was performed, but the operation time was only of seven days.
Sobas \etal \cite{dmma-sm-24} measured the degradation of various \si{\qty{16}{\nano\meter}} FinFET FPGAs with a test bench based on \glspl{ro}.
The devices were monitored as they were aged for \si{\qty{8000}{\hour}}, and the data was applied to develop an empirical degradation model to predict degradation with a relative error of \si{\qty{10}{\percent}}.
Although these measurements have been performed after real-time ageing (\ie not artificially accelerated), the devices were operated under laboratory conditions, which is usually not the case for embedded system deployments.
In this work we provide measurements and analysis of propagation delay of naturally-aged FPGAs, which are currently under operation deployed in the \gls{xfel}.
 \section{Measurement Method} \label{sec:measurement}

\subsection{Procedure Overview}\label{subsec:procedure}
\autoref{fig:procedure} illustrates the steps we follow to perform the propagation delay measurement test.
Given that the proposed test is online, it does not require a relocation of the devices (\eg transportation to a test facility).
The test is performed entirely automatically and remotely, which is extremely important in large deployments like the \gls{xfel}, where human access is impossible during operation.

We start by configuring a test firmware that includes the \gls{pdm} module.
We set the required \gls{ro} quantity, amount of stages, and measurement time per \gls{ro}.
This generates a configuration bitfile that is used to program all operational devices.
Upon a remote trigger via the PCIe connection on the boards, the self-test executes on all devices in parallel.
At the end of each measurement iteration, the results are collected from the RAM registers and the test is re-triggered if necessary.
When all tests are completed, we restore the original firmware to all devices, so they can resume normal operation.
Finally, we train \gls{ml} regressors with the collected measurement results.
We evaluate their performance in estimating the switching frequency of \glspl{ro}, based on operational conditions and radiation exposure.

\begin{figure}[t]
    \centering
    \includegraphics[width=\linewidth]{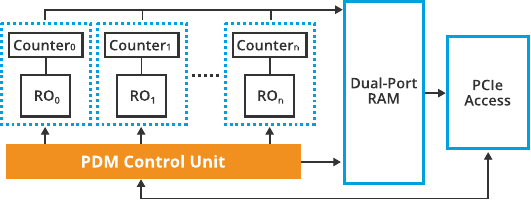}
    \setlength{\belowcaptionskip}{-10pt}
    \setlength{\abovecaptionskip}{-10pt}
    \caption{Propagation delay module consisting of ring oscillators and associated counters, managed by a control unit.
            PCIe registers provide users with access to the control unit and the measurement results.}
    \label{fig:pdm_module}
\end{figure}

\subsection{The Propagation Delay Measurement Module} \label{subsec:pdm}

Due to the high-speed processing and versatility that the control systems at the \gls{xfel} need, most of the deployed electronic boards are based on FPGA devices.
Harsh operating conditions make the collection of diagnostic data from this hardware a strict requirement for analysis, fault prediction, and predictive maintenance.
To cope with the heterogeneity of the deployed devices (\eg different boards and FPGA models), a generic module to measure propagation delay was developed based on the open source DESY FPGA Framework \cite{fwk-bbbd-23}, which provides a convenient hardware abstraction and register-based access.
This allows implementing the module as a library that can be seamlessly integrated to other firmware applications of the accelerator.
The flexibility and re-utilization of the module across different FPGA-based embedded devices are key advantages, as they reduce the development time and the risk of introducing firmware errors.

\autoref{fig:pdm_module} shows the three main parts of the \gls{pdm} module: \one the \glspl{ro} and counters, \two the control unit that manages the \gls{ro} activation, and \three the connectivity via PCIe access registers.
The module allows instantiating an arbitrary and configurable number of \glspl{ro} on the FPGA, each of which is connected to its own oscillations counter.
The number of stages that compose the \glspl{ro} is configurable at design time.
When a new measurement is triggered, the control unit enables the \glspl{ro} one by one for a pre-determined and configurable time.
Between each \gls{ro} measurement, the intermediate results (\ie number of oscillations) are copied into an SRAM memory.
Exposing the module via registers enables performing measurements and reading out the results on demand via PCIe.
PCIe connectivity is particularly important for this type of facilities, as it is a widely utilized bus to interact with hardware and to build control systems \cite{vkhh-ctkca-17}.

In addition to the VHDL part of the module, we developed a placement tool to distribute the \gls{ro} instances along the FPGA fabric at design time.
Designers can indicate the number of required \glspl{ro}, the desired spacing and surface for each (expressed in number of slices), as well as the FPGA chip model.
Optionally, further rules can be applied, namely: a list of forbidden areas (where \glspl{ro} are not to be placed), a list of exclusive areas (\ie a \gls{ro} cannot overlap with more of one of these at a time), and a list of pre-fixed positions where \glspl{ro} must be placed.
Utilizing the provided restrictions, the tool distributes the \glspl{ro} on the available area and outputs a constraints file that can be added to the firmware project to fix the instances during the place and route process.
This file also ensures that all \glspl{ro} share the same structure by defining the logic placement at LUT level.

\subsection{Radiation Environment at the European XFEL} \label{subsec:radiation}

The environments of linear particle accelerators present mixed neutron and gamma radiations during operation.
The three major sources of radiation are:
\one field radiation due to the presence of areas with high electric fields, which are induced by imperfections at the atomic level on the inner parts of the RF cavities,
\two accelerated electrons from the RF gun (\ie dark current) that impact on materials causing gamma radiation, and
\three bremsstrahlung due to accelerated charged particles \cite{rmsx-mmsj-07}.

Due to the enhanced background radiation, electronic devices are typically placed outside main accelerator tunnels. This is not the case for the \gls{xfel}.
Given the low-latency requirements of the \gls{llrf} and other systems, the electronic equipment has been placed inside the tunnel near the accelerator pipe.
To attempt reducing radiation doses, devices are positioned underneath a \si{\qty{20}{\centi\meter}} thick concrete cover.
Although this reduces the radiation exposure, it does not shield completely from neutrons, secondary, and tertiary showers.

To explore possible correlations between propagation delay and radiation exposure, we utilize radiation dose data gathered weekly from the \gls{xfel} tunnel by a mobile autonomous robot: MARWIN \cite{marwin-dmh-17}.
The robot is capable of moving along the tunnel while measuring gamma and neutron instant radiation dose rates at each point.
Thus, providing a good approximation of the current radiation conditions on which the \gls{llrf} electronics operate.

\subsection{Training of Machine Learning Models}\label{subsec:eval_ml}

Dependable and mission-critical systems such as the \gls{llrf} system at the \gls{xfel} cannot afford unscheduled downtime.
To avoid such eventualities, it is crucial to correctly estimate when components are due for replacement before their performance degrades beyond acceptable limits and system faults appear.
Consequently, it is paramount to implement constant monitoring and predictive maintenance with estimator models.

The degradation of the propagation delay is a critical issue for FPGA-based applications in the accelerator.
Therefore, we evaluate the performance of multiple \gls{ml} models in the estimation of ring oscillator frequencies, given historical and environmental data.
For all models the input features are: FPGA die temperature and core voltage at the moment of the frequency measurement, gamma and neutron radiation doses to which the devices have been exposed, the device location in the tunnel, the position in the MicroTCA crate, and the \gls{ro} location on the chip.

In this work, we evaluate seven regressor \gls{ml} models: six ensemble methods, and a linear model fitted using the \gls{sgd} algorithm.
The main idea behind ensemble models is to train and combine various base models in a way that together they achieve better results.
The major strategies to combine and train the base models are boosting, bagging, and stacking.
The ensemble models evaluated here are based on trees: three of them are boosting models (histogram-based gradient boosting \cite{sgb-f-02}, XGBoost \cite{xgb-cg-16}, and AdaBoost \cite{adab-fs-97}), and three are bagging models (simple bagging regressor \cite{bagg-b-96}, random forest \cite{rf-b-01}, and extremely randomized trees \cite{ert-gew-06}).
We utilize implementation of XGBoost provided by the authors, while for the other models we make use of the implementations from the Scikit-learn library \cite{scikitlearn}.

For the training process, we randomly split the measured operational devices into a training group (\si{\qty{70}{\percent}}) and a test group (\si{\qty{30}{\percent}}), so that models are tested on measurements from devices which are unseen during training.
We then perform a hyperparameter tuning process, where we train and evaluate the models under different settings to find the best performant configuration.

In order to prevent leakage of the test dataset into the training process when evaluating the different settings, we apply a cross validation method known as group k-fold.
With this method, the evaluation during hyperparameter tuning is performed using a subset of the training dataset instead of the test dataset.
The method consists in splitting the training dataset into \(k\) subsets (or folds) while making sure that no measurements from the same device belong to two different folds.
The model is then trained and evaluated \(k\) times for each hyperparameter set.
Each time, \(k-1\) folds are used for training, and the remaining fold is used for evaluation.
Every fold is used once as evaluation set during tuning.
The average performance of the model over all iterations is used to select the best performant hyperparameter set.
Finally, the best model is evaluated on the unseen test dataset.
 
\begin{figure}[!t]
    \centering
    \includegraphics[width=\linewidth]{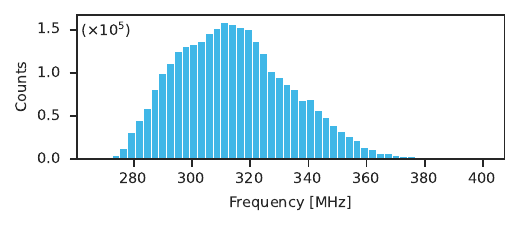}
    \setlength{\belowcaptionskip}{-10pt}
    \setlength{\abovecaptionskip}{-10pt}
    \caption{Frequency distribution the ring oscillator measurements on operational devices.}
    \label{fig:all_ro_histogram}
\end{figure}

\section{Measurement Results} \label{sec:results}

In this section we present and compare the results of the propagation delay measurements from used and unused devices.
We analyse the influence radiation dose rates, and measure the performance of \gls{ml} models in frequency estimation.

\subsection{Hardware Under Study} \label{subsec:studied_hardware}

This study focuses on FPGA-based boards that belong to the \gls{llrf} system of the \gls{xfel}.
In particular, we examine digitizer boards of this system, which feature Xilinx Virtex-6 XC6VLX130T (speed grade -2) FPGAs.
All devices under study shared the same firmware and usage during operation.

In our experiments, we evaluate \si{\num{298}} devices that have been in operation since 2017 inside the accelerator tunnel.
We conduct the measurements during the \gls{xfel} winter shutdown period of 2023, so that machine operation is not affected.
The test firmware utilized in our experiments instantiates \si{\num{100}} \glspl{ro} on the FPGAs, providing a good resolution for the propagation delay measurements.
Each of the \glspl{ro} is active (\ie oscillates) for \si{\qty{500}{\milli\second}} at a time, and the number of oscillations is temporally stored in an on-chip SRAM to be read out via PCIe registers.
We perform \si{\num{100}} iterations of the propagation delay measurement per device, and we register the FPGA die temperature and core voltage each time utilizing onboard sensors.

To create a baseline for comparison with the deployed devices under study, we perform the exact same measurements on \si{\num{7}} unused devices.
These devices have never been deployed and have been kept in storage as replacement parts of the \gls{llrf} system.
Given that they have not been exposed to high radiation doses nor powered during storage, no significant NBTI, HCI, nor TDI effects have been at play.

\subsection{Frequency Distribution} \label{subsec:frequency_distribution}

We measure all \glspl{ro} from each device in operation \si{\num{100}} times, which yields a total of \si{\num{2960000}} measurements.
\autoref{fig:all_ro_histogram} depicts a histogram with the distribution of these measurements.
We observe that the mean oscillation frequency sits at \si{\qty{313}{\mega\hertz}}, while the total distribution spans from \si{\qty{270}{\mega\hertz}} up to \si{\qty{400}{\mega\hertz}}.
The distribution of frequencies is positively skewed, and it displays a distance to the mean frequency of around \si{\qty{25}{\percent}} for the higher values and \si{\qty{16}{\percent}} for the lower values.
This skewness could be related to the way speed grades are measured by FPGA manufacturers: to classify a device as a certain speed grade, only minimum requirements are set.
Therefore, it is possible that two devices with substantially different propagation delay characteristics belong to the same speed grade, as long as they meet these requirements and regardless of the difference among them.

\begin{figure}[!t]
    \centering
    \includegraphics[width=\linewidth]{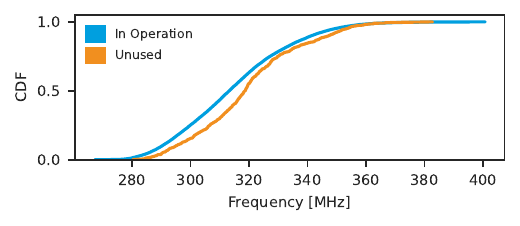}
    \setlength{\belowcaptionskip}{-15pt}
    \setlength{\abovecaptionskip}{-15pt}
    \caption{Cumulative distribution functions of the frequency measurements from used and unused devices.}
    \label{fig:cumulative_distribution}
\end{figure}

\subsection{Comparison with Unused Devices}\label{subsec:unused_devices}

Given that no measurements exist of the used devices prior to deployment, we utilize a group of seven boards kept in storage as a comparison baseline.
They are composed of the exact same hardware, and the test configuration is also equal to the one applied to used boards.

\begin{figure*}[t]
    \centering
    \begin{subfigure}[t]{\linewidth}
        \centering
        \includegraphics[width=\linewidth]{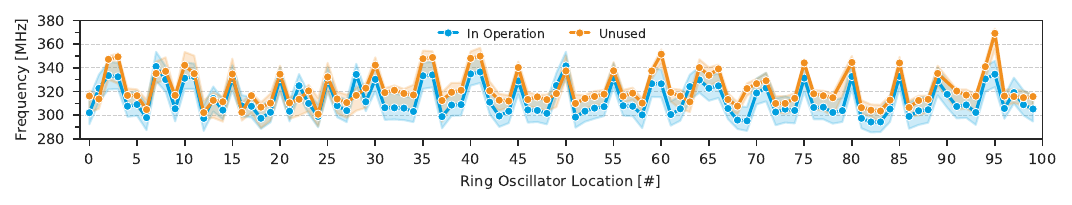}
        \caption{Median frequency measured on used and unused devices.
        }
        \label{fig:frequency_by_index}
    \end{subfigure}

    \hfill

    \begin{subfigure}[t]{\linewidth}
        \centering
        \includegraphics[width=\linewidth]{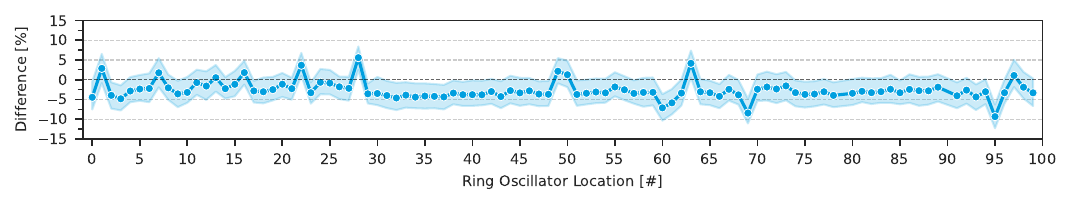}
        \caption{Median frequency difference of used devices relative to the median frequency of unused devices for the same location.
        }
        \label{fig:unused_frequency_diff_by_index}
    \end{subfigure}

    \caption{Comparison of frequencies between used and unused devices for each \gls{ro} location. The error bands represent the interquartile ranges of the values. }
\end{figure*}

\autoref{fig:cumulative_distribution} shows the cumulative distribution functions for used and unused devices.
For most of the lower frequencies, we observe that the distribution of operational devices stays above the unused ones.
Moreover, the frequency span on unused devices is slightly smaller than the one of devices in operation, as the latter group reaches frequencies lower than \si{\qty{280}{\mega\hertz}}.
The distribution functions present a particularly larger difference on frequencies below \si{\qty{320}{\mega\hertz}} and present a crossing only slightly below \si{\qty{360}{\mega\hertz}}.
This indicates that the frequency values for unused devices concentrate mainly towards higher frequencies, in comparison to operational devices.
Indeed, as the unused devices have not been exposed to strong degradation mechanisms, their propagation delay times are not affected and tend to yield faster oscillation frequencies.

To statistically compare the distributions of frequencies from used and unused devices, we employ the Brunner-Munzel test \cite{npbfp-bm-00} (otherwise known as the generalized Wilcoxon test).
This is a non-parametric test for stochastic equality of samples belonging to two independent groups.
The test is well suited to analyse groups with at least ten samples, which makes it applicable in our case due to the number of measurements points available per device.
The null hypothesis \(H_0\) of this test states that when picking values from each group there are equal probabilities of getting large values from both groups, while the alternative hypothesis is that one group (the unused devices in this case) has larger values than the other.
The reasons to select this test are twofold:
\one given that the measurements show skewness, we use a non-parametric test that does not assume an underlying standard distribution,
and \two it does not require an equal variance between both sample groups.
The test statistic \(B\) is calculated as in \autoref{eq:brunner_munzel}, where \(\overline{R}_1\) and \(\overline{R}_2\) are the means of the midranks from both groups, \(n_1\) and \(n_2\) are the number of observations on each group, and \(\sigma^2_1\) and \(\sigma^2_2\) are the variance estimates of the groups.

\begin{equation}
    B = \frac{\overline{R}_2 - \overline{R}_1}{(n_1 + n_2) \sqrt{\frac{{\sigma^2_1}}{n_1} + \frac{{\sigma^2_2}}{n_2}} }
    \label{eq:brunner_munzel}
\end{equation}

To reject \(H_0\) of the Brunner-Munzel test, we consider a significance value \(\alpha=1\%\), meaning that the \emph{p} value calculated from the test statistic should be lower than \si{\num{0.01}} in order to reject the null hypothesis with sufficient confidence.
We utilize the implementation from the SciPy package \cite{scipy} to compute the test, which yields a statistic value of \si{\num{7.194}} and a \emph{p} value of \si{\num{7.827e-13}} calculated using a t-distribution.
Considering the rejection criteria, we can assume \(H_0\) to be false within \si{\qty{1}{\percent}}.
Therefore, we can accept the alternative hypothesis that the frequencies from the unused devices group are significantly higher than the used ones.

By breaking down the frequency comparison between used and unused devices over \gls{ro} locations, we can reduce the noise of inherent frequency differences introduced by the die area and circuit routing.
We are able to compare \glspl{ro} on the same die location across devices because the same test configuration (\ie the exact same bitfile) has been used for all measurements.
\autoref{fig:frequency_by_index} shows the median frequency on each \gls{ro} location for unused and operational devices, with a shaded area depicting the interquartile range (IQR).
Measurements for unused devices on locations 79 and 90 are not be included due to data corruption.
We can see that a pattern of frequencies emerges across locations on both cases, indicating that there is a dependence between the location of the \gls{ro} and its measured frequency.
Given that the patterns on used and unused devices are similar, we cannot attribute them solely to transistor ageing on that die area.
Delay differences introduced by \gls{ro} signal routing is most likely contributing to this location dependence.
Except for some location outliers, we observe that on most of the \glspl{ro} the median oscillation frequency is higher for unused devices than for devices in operation.
In some cases the median of unused devices is even outside the marked IQR for operational devices.

Let \(f^u(l)\) be all frequency measurements from unused devices at the \gls{ro} location \(l\), and let \(\widetilde{f}^u(l)\) be their median value.
\(f^d_i(l)\) represents the frequency measurement number \(i\) from the device \(d\) at the same location \(l\).
We calculate \(\Delta^d_i(l)\) as expressed in \autoref{eq:freq_diff}, which is the frequency difference relative to \(\widetilde{f}^u(l)\).

\begin{equation}
    \Delta^d_i(l) = \frac{f^d_i(l) - \widetilde{f}^u(l)}{\widetilde{f}^u(l)} \times 100
    \label{eq:freq_diff}
\end{equation}

\autoref{fig:unused_frequency_diff_by_index} depicts for each \gls{ro} location the median value of \(\Delta\) as dots and the IQR as a shaded area around them.
On most locations, we observe a negative median difference except for some outlier \glspl{ro}.
The tendency seems to be uniform across locations, with values mostly ranging between \si{\qty{3}{\percent}} and \si{\qty{5}{\percent}}.
This is in line with the previous observation that unused devices present higher frequencies (\ie lower propagation delays) than deployed devices, due to the presence of degradation mechanisms in the latter.

\begin{figure*}[!t]
    \centering
    \begin{subfigure}[t]{0.49\linewidth}
        \centering
        \includegraphics[width=\linewidth]{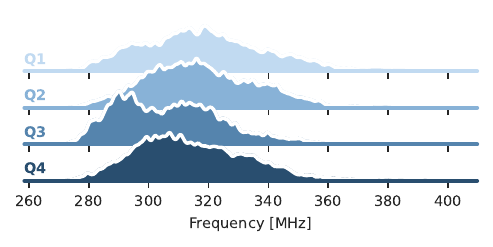}
        \setlength{\belowcaptionskip}{-3pt}
        \setlength{\abovecaptionskip}{-3pt}
        \caption{Frequencies for gamma dose rate quartiles.}
        \label{fig:gamma_quartiles}
    \end{subfigure}
    \hfill
    \begin{subfigure}[t]{0.49\linewidth}
        \centering
        \includegraphics[width=\linewidth]{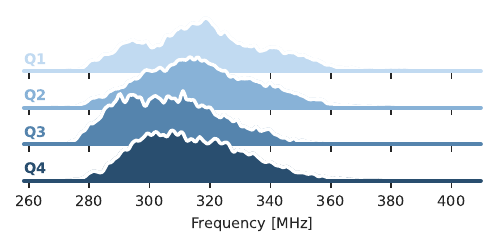}
        \setlength{\belowcaptionskip}{-3pt}
        \setlength{\abovecaptionskip}{-3pt}
        \caption{Frequencies for neutron dose rate quartiles.}
        \label{fig:neutron_quartiles}
    \end{subfigure}

    \caption{Kernel density estimations of \gls{ro} frequency distributions for devices on all radiation dose rate quartiles, where Q1 contains the \si{\qty{25}{\percent}} of devices with the least radiation and Q4 the \si{\qty{25}{\percent}} with the most.}
    \label{fig:radiation_kdes}
\end{figure*}

\subsection{Influence of Gamma and Neutron Radiation}\label{subsec:effect_radiation}

We now analyse whether the \gls{ro} frequency measurements show any correlations to the radiation doses to which the FPGAs have been exposed during operation.
To this end, we utilize the radiation dose rate data gathered by the autonomous robot in the \gls{xfel} tunnel.
Radiation measurements are performed weekly, and sensor readings occur at every point of the way along the \si{\qty{1.7}{\kilo\meter}} of route.
To assign a concrete dose rate value to a \gls{llrf} crate, we average the radiation values read in the vicinity of \si{\qty{+- 5}{\meter}} of the electronic devices.
Moreover, we consider that devices belonging to a given crate receive similar radiation dose rates.
As radiation levels inside the accelerator do not present large variations over time, calculating the average dose rates from the weekly readings is a good estimation of the exposure on each device.

We form radiation quartiles by dividing the population of devices into four groups according to their average dose rates, so that each group contains \si{\qty{\approx 25}{\percent}} of the devices.
We now compare the \gls{ro} frequency distribution across devices with increasing gamma and neutron dose rates, as depicted in the ridge plots of \autoref{fig:radiation_kdes}.
On both radiation types, we observe a strong trend towards frequency reduction for the quartiles Q1, Q2, and Q3, that is, as radiation dose rates increase.
The \si{\qty{25}{\percent}} of devices on the higher radiation part of the population (\ie Q4) do not seem to follow the exact same pattern when compared with Q3, but still present a tendency towards lower frequencies in comparison to Q2 and Q1.
A possible explanation for the outlying behaviour is an inherent difference already present in the initial frequency distribution.
In the case that more devices in Q4 were originally faster than the ones in Q3, even after undergoing degradation this difference could affect the quartiles distributions.
Unfortunately, the information of the deployed devices before ageing is not available.
The observed trend across quartiles indicates that there is a tendency towards lower frequencies as radiation dose rates of devices increase, but the trend is not fully continuous.
This correlation between dose rates and frequency reduction indicates possible effects of the TID degradation mechanism on the operational devices as presented in \autoref{sec:introduction}.

\subsection{Ring Oscillator Frequency Estimation}\label{subsec:ml}

We employ two metrics to evaluate the performance of the different \gls{ml} models: \gls{mape}, and coefficient of determination (\ie \(R^2\) score).
The \gls{mape} provides an interpretable value of the inference error and is independent of the actual magnitude.
The \(R^2\) score provides a comparison with a na\"ive baseline model that always outputs the mean value of the data points and that scores \(R^2 = 0\).
The score is calculated as

\begin{equation*}
    R^2(y, \hat{y}) = 1 - \frac{\sum_{i=1}^{n}(y_i - \hat{y}_i)^2}{\sum_{i=1}^{n}(y_i - \bar{y})^2}
\end{equation*}

where \(y_i\) is the true \(i\)-th value, \(\hat{y}_i\) is the model prediction, \(\bar{y}\) is the average of the ground truth values, and \(n\) is the total number of values.
A model can get a maximum \(R^2\) of 1 (\ie best score) as well as negative values.

\begin{figure}[t]
    \centering
    \includegraphics[width=\linewidth]{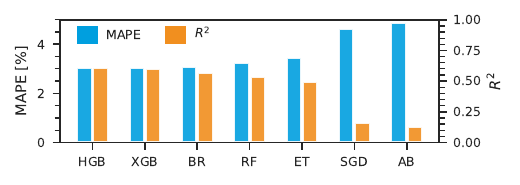}
    \caption{Mean absolute percentage error and \(R^2\) score of the evaluated models: Histogram-based Gradient Boosting (HGB), XGBoost (XGB), Simple Bagging Regressor (BR), Random Forest (RF), Extra Trees (ET), linear fitted with \gls{sgd}, and AdaBoost (AB).}
    \label{fig:regressor_performance}
\end{figure}

\autoref{fig:regressor_performance} shows the \glspl{mape} and the \(R^2\) scores for each evaluated model, sorted from left to right in increasing \gls{mape} value.
We observe that the error ranges between \si{\qty{3.02}{\percent}} and \si{\qty{4.85}{\percent}}.
The HGB and XGBoost models perform the best with \(R^2\) scores around \si{\num{0.6}} and \glspl{mape} close to \si{\qty{3.02}{\percent}}.
On the opposite end, the AdaBoost model shows a poor performance in terms of \gls{mape} (\si{\qty{4.85}{\percent}}) and \(R^2\) (\si{\num{0.125}}), similar to the linear model.
Except for AdaBoost, boosting ensemble models (gradient boosting and XGBoost) perform slightly better than the bagging ensemble models (simple bagging, random forest, and extra trees), followed by the linear model fitted with \gls{sgd}.

 \section{Conclusions and Future Work} \label{sec:conclusions_future_work}

In this work, we presented a \gls{ro}-based module to measure propagation delay on FPGAs, as well as a procedure to execute the module as an online self-test.
We deployed the module and measured 298 naturally-aged FPGA devices that are currently in operation at the \gls{xfel}.
Based on our measurements, we studied the influence of ageing mechanisms on the propagation delay of the devices.
We found that FPGAs in operation present slower switching frequencies with statistical significance when compared with unused devices, due to the effects of hardware degradation such as BTI and TDI.
Moreover, we identified a correlation between increased delays and higher radiation doses, suggesting that radiation has had a negative influence in the performance of the deployed FPGA devices.

Additionally, we have shown the feasibility of \gls{ml} models that estimate the propagation delay of \glspl{ro} based on historical and environmental data.
We trained and evaluated various regressor models that yielded mean absolute percentage errors as low as \si{\qty{3.02}{\percent}} and \(R^2\) scores as high as \si{\num{0.608}}.

In the future, we plan to extend the \gls{pdm} module for online concurrent self-tests that can be deployed together with the application and collect data while the accelerator is in operation.
With more data at different points in time, we aim to better understand the time-dependence of the degradation effects on FPGAs deployed in large-scale on a highly-dependable and harsh environment.
Moreover, we intend to evaluate the performance of \gls{ml} models when predicting propagation delay degradation based on time-series data, which can allow developing early warning systems for predictive maintenance.
 
\ifanonymous
\else
\section*{Acknowledgement}
We acknowledge the support by DASHH (Data Science in Hamburg - HELMHOLTZ Graduate School for the Structure of Matter) with the Grant-No. HIDSS-0002,
as well as the support of the Federal Ministry of Education and Research with Grant C-ray4edge.

 \fi

\bibliographystyle{IEEEtran}

\begin{thebibliography}{10}
\providecommand{\url}[1]{#1}
\csname url@samestyle\endcsname
\providecommand{\newblock}{\relax}
\providecommand{\bibinfo}[2]{#2}
\providecommand{\BIBentrySTDinterwordspacing}{\spaceskip=0pt\relax}
\providecommand{\BIBentryALTinterwordstretchfactor}{4}
\providecommand{\BIBentryALTinterwordspacing}{\spaceskip=\fontdimen2\font plus
\BIBentryALTinterwordstretchfactor\fontdimen3\font minus \fontdimen4\font\relax}
\providecommand{\BIBforeignlanguage}[2]{{\expandafter\ifx\csname l@#1\endcsname\relax
\typeout{** WARNING: IEEEtran.bst: No hyphenation pattern has been}\typeout{** loaded for the language `#1'. Using the pattern for}\typeout{** the default language instead.}\else
\language=\csname l@#1\endcsname
\fi
#2}}
\providecommand{\BIBdecl}{\relax}
\BIBdecl

\bibitem{xfel-06}
R.~Abela, A.~Aghababyan, M.~Altarelli, C.~Altucci, and G.~Amatuni, \emph{{XFEL: The European X-Ray Free-Electron Laser - Technical Design Report}}.\hskip 1em plus 0.5em minus 0.4em\relax Hamburg: DESY, 2006.

\bibitem{llrf-baag-12}
J.~Branlard, G.~Ayvazyan, V.~Ayvazyan, M.~Grecki, M.~Hoffmann, F.~Ludwig, U.~Mavri\v{c}, S.~Pfeiffer, C.~Schmidt, H.~Weddig, B.~Yang, H.~Schlarb, S.~Bou~Habib, L.~Butkowski, K.~Czuba, M.~Grzegrzolka, E.~Janas, J.~Piekarski, and W.~Cichalewski, ``{The European XFEL LLRF System},'' in \emph{Proceedings of the 3rd International Particle Accelerator Conference}, New Orleans, Louisiana, USA, 2012.

\bibitem{lmfssw-tadmes-23}
\BIBentryALTinterwordspacing
L.~Lanzieri, G.~Martino, G.~Fey, H.~Schlarb, T.~C. Schmidt, and M.~W{\"a}hlisch, ``A review of techniques for ageing detection and monitoring on embedded systems,'' 2023. [Online]. Available: \url{https://arxiv.org/pdf/2301.06804}
\BIBentrySTDinterwordspacing

\bibitem{nbti-s-07}
D.~K. Schroder, ``Negative bias temperature instability: What do we understand?'' \emph{Microelectronics Reliability}, vol.~47, no.~6, pp. 841--852, 2007, modelling the Negative Bias Temperature Instability.

\bibitem{rahd-swz-10}
J.~Stathis, M.~Wang, and K.~Zhao, ``Reliability of advanced high-k/metal-gate {n-FET} devices,'' \emph{Microelectronics Reliability}, vol.~50, no.~9, pp. 1199--1202, 2010, 21st European Symposium on the Reliability of Electron Devices, Failure Physics and Analysis.

\bibitem{tidc-b-06}
H.~J. Barnaby, ``Total-ionizing-dose effects in modern {CMOS} technologies,'' \emph{IEEE Transactions on Nuclear Science}, vol.~53, no.~6, pp. 3103--3121, 2006.

\bibitem{lkfss-aaesl-23}
L.~Lanzieri, P.~Kietzmann, G.~Fey, H.~Schlarb, and T.~C. Schmidt, ``{Ageing Analysis of Embedded SRAM on a Large-Scale Testbed Using Machine Learning},'' in \emph{Proc. of 26th Euromicro Conference on Digital System Design (DSD)}.\hskip 1em plus 0.5em minus 0.4em\relax Piscataway, NJ, USA: IEEE, September 2023, pp. 335--342.

\bibitem{oshf-zh-10}
K.~M. Zick and J.~P. Hayes, ``On-line sensing for healthier {FPGA} systems,'' in \emph{Proceedings of the 18th Annual ACM/SIGDA International Symposium on Field Programmable Gate Arrays}, ser. FPGA '10.\hskip 1em plus 0.5em minus 0.4em\relax New York, NY, USA: ACM, 2010, pp. 239--248.

\bibitem{dmma-sm-24}
J.~Sobas and F.~Marc, ``{Degradation Measurement and Modelling under Ageing in a 16 nm FinFET FPGA},'' \emph{Micromachines}, vol.~15, no.~1, p.~19, 2024.

\bibitem{ispr-vs-19}
V.~A. Chhabria and S.~S. Sapatnekar, ``{Impact of Self-heating on Performance and Reliability in FinFET and GAAFET Designs},'' in \emph{20th International Symposium on Quality Electronic Design (ISQED)}.\hskip 1em plus 0.5em minus 0.4em\relax Piscataway, NJ, USA: IEEE, 2019, pp. 235--240.

\bibitem{gdrm-os-95}
S.~Ogawa and N.~Shiono, ``Generalized diffusion-reaction model for the low-field charge-buildup instability at the si-${\mathrm{sio}}_{2}$ interface,'' \emph{Phys. Rev. B}, vol.~51, pp. 4218--4230, Feb 1995.

\bibitem{rwmc-swvs-09}
A.~W. Strong, E.~Y. Wu, R.-P. Vollertsen, J.~Suñe, G.~La~Rosa, S.~E. Rauch, and T.~D. Sullivan, ``Hot carriers,'' in \emph{Reliability Wearout Mechanisms in Advanced CMOS Technologies}.\hskip 1em plus 0.5em minus 0.4em\relax Hoboken, NJ, USA: John Wiley \& Sons, Ltd, 2009, ch.~5, pp. 441--516.

\bibitem{hcff-jlkk-16}
M.~Jin, C.~Liu, J.~Kim, J.~Kim, S.~Choo, Y.~Kim, H.~Shim, L.~Zhang, K.-j. Nam, J.~Park, S.~Pae, and H.~Lee, ``Hot carrier reliability characterization in consideration of self-heating in {FinFET} technology,'' in \emph{2016 IEEE International Reliability Physics Symposium (IRPS)}.\hskip 1em plus 0.5em minus 0.4em\relax Piscataway, NJ, USA: IEEE, 2016, pp. 2A--2--1--2A--2--5.

\bibitem{iact-rlnk-17}
R.~Ranjan, Y.~Liu, T.~Nigam, A.~Kerber, and B.~Parameshwaran, ``{Impact of AC voltage stress on core NMOSFETs TDDB in FinFET and planar technologies},'' in \emph{2017 IEEE International Reliability Physics Symposium (IRPS)}.\hskip 1em plus 0.5em minus 0.4em\relax Piscataway, NJ, USA: IEEE, 2017, pp. DG--10.1--DG--10.5.

\bibitem{stro-sbwm-13}
G.~J. Schlenvogt, H.~J. Barnaby, J.~Wilkinson, S.~Morrison, and L.~Tyler, ``Simulation of {TID} effects in a high voltage ring oscillator,'' \emph{IEEE Transactions on Nuclear Science}, vol.~60, no.~6, pp. 4547--4554, 2013.

\bibitem{iama-lhww-22}
Z.~Li, Z.~Huang, Q.~Wang, J.~Wang, and N.~Luo, ``{Implementation of Aging Mechanism Analysis and Prediction for XILINX 7-Series FPGAs with a 28-nm Process},'' \emph{Sensors}, vol.~22, no.~12, p. 4439, 2022.

\bibitem{rloc-pkp-14}
P.~Pfeifer, B.~Kaczer, and Z.~Pliva, ``A reliability lab-on-chip using programmable arrays,'' in \emph{2014 IEEE International Reliability Physics Symposium}.\hskip 1em plus 0.5em minus 0.4em\relax Piscataway, NJ, USA: IEEE, 2014, pp. CA.6.1--CA.6.8.

\bibitem{fwk-bbbd-23}
L.~Butkowski, A.~Bellandi, M.~B\"{u}chler, B.~Dursun, C.~G\"{u}m\"{u}\c{s}, N.~Omidsajedi, and K.~Schulz, ``{The DESY Open Source FPGA Framework},'' in \emph{Proceedings of the 19th International Conference on Accelerator and Large Experimental Control Systems}, Cape Town, South Africa, 2023.

\bibitem{vkhh-ctkca-17}
G.~Varghese, M.~Killenberg, M.~Heuer, M.~Hierholzer, L.~Petrosyan, C.~Schmidt, A.~Piotrowski, A.~Dworzanski, K.~Czuba, C.~Iatrou \emph{et~al.}, ``Chimeratk-a software tool kit for control applications,'' in \emph{IPAC17, Copenhagen, Denmark}, 2017.

\bibitem{rmsx-mmsj-07}
D.~Makowski, B.~Mukherjee, S.~Simrock, G.~Jabłoński, A.~Napieralski, and M.~Grecki, ``{Radiation monitoring system for XFEL},'' \emph{Measurement Science and Technology}, vol.~18, no.~8, p. 2397, jul 2007.

\bibitem{marwin-dmh-17}
A.~Dehne, N.~Moller, and T.~Hermes, ``{MARWIN}: a mobile autonomous robot for maintenance and inspection in a 4d environment,'' in \emph{2017 International Conference on Research and Education in Mechatronics (REM)}, 2017, pp. 1--5.

\bibitem{sgb-f-02}
J.~H. Friedman, ``Stochastic gradient boosting,'' \emph{Computational Statistics \& Data Analysis}, vol.~38, no.~4, pp. 367--378, 2002, nonlinear Methods and Data Mining.

\bibitem{xgb-cg-16}
T.~Chen and C.~Guestrin, ``Xgboost: A scalable tree boosting system,'' in \emph{Proceedings of the 22nd acm sigkdd international conference on knowledge discovery and data mining}, 2016, pp. 785--794.

\bibitem{adab-fs-97}
Y.~Freund and R.~E. Schapire, ``A decision-theoretic generalization of on-line learning and an application to boosting,'' \emph{Journal of Computer and System Sciences}, vol.~55, no.~1, pp. 119--139, 1997.

\bibitem{bagg-b-96}
L.~Breiman, ``Bagging predictors,'' \emph{Machine learning}, vol.~24, pp. 123--140, 1996.

\bibitem{rf-b-01}
L.~Breiman, ``Random forests,'' \emph{Machine learning}, vol.~45, pp. 5--32, 2001.

\bibitem{ert-gew-06}
P.~Geurts, D.~Ernst, and L.~Wehenkel, ``Extremely randomized trees,'' \emph{Machine learning}, vol.~63, pp. 3--42, 2006.

\bibitem{scikitlearn}
F.~Pedregosa, G.~Varoquaux, A.~Gramfort, V.~Michel, B.~Thirion, O.~Grisel, M.~Blondel, P.~Prettenhofer, R.~Weiss, V.~Dubourg, J.~Vanderplas, A.~Passos, D.~Cournapeau, M.~Brucher, M.~Perrot, and E.~Duchesnay, ``Scikit-learn: Machine learning in {P}ython,'' \emph{Journal of Machine Learning Research}, vol.~12, pp. 2825--2830, 2011.

\bibitem{npbfp-bm-00}
E.~Brunner and U.~Munzel, ``The nonparametric behrens-fisher problem: asymptotic theory and a small-sample approximation,'' \emph{Biometrical Journal: Journal of Mathematical Methods in Biosciences}, vol.~42, no.~1, pp. 17--25, 2000.

\bibitem{scipy}
P.~Virtanen, R.~Gommers, T.~E. Oliphant, M.~Haberland, T.~Reddy, D.~Cournapeau \emph{et~al.}, ``{{SciPy} 1.0: Fundamental Algorithms for Scientific Computing in Python},'' \emph{Nature Methods}, vol.~17, pp. 261--272, 2020.

\end{thebibliography}

\balance

\label{lastpage}

\end{document}